\documentclass[aps,prl,showpacs,twocolumn, superscriptaddress]{revtex4}%

\newcommand{\ket}[1]{|#1\rangle}
\newcommand{\braket}[2]{\langle #1|#2\rangle}

\usepackage{amsfonts}
\usepackage{amsmath}
\usepackage{amssymb}
\usepackage{graphicx}

\begin{document}

\title{Imaging the symmetry breaking of molecular orbitals in carbon nanotubes}

\author{H. Lin}
\affiliation{Laboratoire Mat\'{e}riaux et Ph\'{e}nom\`{e}nes Quantiques, Universit\'{e} Paris Diderot-CNRS, UMR 7162, 10 rue Alice Domon et L\'{e}onie Duquet, 75205 Paris Cedex 13, France}%
\affiliation{Laboratoire d'Etude des Microstructures,
ONERA-CNRS, BP72, 92322 Ch\^{a}tillon,
France}
\author{J. Lagoute}
\author{V. Repain}
\author{C. Chacon}
\author{Y. Girard}
\affiliation{Laboratoire Mat\'{e}riaux et Ph\'{e}nom\`{e}nes Quantiques, Universit\'{e} Paris Diderot-CNRS, UMR 7162, 10 rue Alice Domon et L\'{e}onie Duquet, 75205 Paris Cedex 13, France}%
\author{F. Ducastelle}
\author{H. Amara}
\author{A. Loiseau}
\affiliation{Laboratoire d'Etude des Microstructures, ONERA-CNRS, BP72, 92322 Ch\^{a}tillon,
France}%
\author{P. Hermet}
\author{L. Henrard}
\affiliation{Research Centre in Physics of Matter and Radiation (PMR), University of Namur (FUNDP), rue de Bruxelles 61, 5000 Namur, Belgium}%
\author{S.~Rousset}
\affiliation{Laboratoire Mat\'{e}riaux et Ph\'{e}nom\`{e}nes Quantiques, Universit\'{e} Paris Diderot-CNRS, UMR 7162, 10 rue Alice Domon et L\'{e}onie Duquet, 75205 Paris Cedex 13, France}%
\date{\today}

\begin{abstract}
Carbon nanotubes have attracted considerable interest for
their unique electronic properties. They are fascinating candidates
for fundamental studies of one dimensional materials as well as for future molecular electronics applications.
The molecular orbitals of nanotubes are of particular importance as they govern the transport properties
and the chemical reactivity of the system. Here we show for the first time a complete experimental investigation of
molecular orbitals of single wall carbon nanotubes using atomically resolved scanning tunneling
spectroscopy. Local conductance measurements show spectacular carbon-carbon bond asymmetry at
the Van Hove singularities for both semiconducting and metallic tubes,
demonstrating the symmetry breaking of molecular orbitals in nanotubes. Whatever the tube, only two types of
complementary orbitals are alternatively observed. An analytical tight-binding model describing the interference patterns of $\pi$ orbitals confirmed by ab initio calculations, perfectly reproduces the experimental results.

\end{abstract}

\pacs{78.67.Ch,73.22.Dj,68.37Ef}

\maketitle

Single-walled carbon nanotubes (SWNTs) are fascinating candidates for
fundamental investigations of electron transport in one-dimensional
systems, as well as for molecular electronics
\cite{NT1,Loiseau2006,Charlier2007}. The control of their electronic bandstructure using doping or functionalization
is a major challenge for future applications.
In this context, the molecular orbitals of SWNTs play a major role since they are fully involved in the transport properties
and the chemical reactivity of SWNTs \cite{Strano2003}. In particular their spatial electronic distribution
is a key factor in chemical reactions \cite{Joselevich2004} and in adsorption process \cite{Wu2004}.
In transport experiments, it is an important parameter for the contact
quality between a tube and an electrode \cite{Tersoff1999}.
A complete understanding and control of the molecular orbitals of nanotubes 
is therefore crucial for the  development of future nanotubes based applications.

Scanning tunneling microscopy and spectroscopy (STM/STS) are unique tools 
to measure local electronic properties of SWNTs and correlate them with their atomic structure.
STS studies have confirmed the predictions of the simple zone folding tight-binding
model which relates the metallic or semiconductor character 
to the chirality of SWNTs \cite{Wildoer1998,Odom1998,Venema2000}. In addition, an electronic ``pseudogap'' on metallic tubes 
has been measured as predicted theoretically when curvature effects or intertube interactions are fully
considered \cite{Ouyang2001}. 
Interference patterns have been
observed in STM images close to defects such as chemical
impurities, cap ends or in finite length nanotubes
\cite{Clauss1999,Furuhashi2008,Venema1999, Lemay2001}.  These
observations have been explained qualitatively from interference
effects between the Bloch waves
\cite{Kane1999,Rubio1999,Furuhashi2008}.
However, the molecular orbitals corresponding to the Bloch states associated to the Van Hove singularities (VHS) have been less investigated.
It has been predicted that, for defect-free
semiconducting tubes, the electronic states would display broken
symmetry effects in agreement with STM topographic images
\cite{Clauss1999,Kane1999,Lambin2003}. The
specific signatures of such Bloch states in
metallic and semiconducting tubes can only be evidenced using local
differential conductance spectroscopy, but experimental data are still
scarce.  

Here, we report on a systematic study of the local densities
of states (LDOS) in defect-free SWNTs at various VHS using a low
temperature STM.  For semiconducting tubes, direct images of the
two first highest occupied molecular-orbitals (HOMO-1, HOMO) and lowest unoccupied molecular-orbitals (LUMO, LUMO+1) states have been
obtained. A spectacular C-C bond asymmetry is observed, revealing the
complementarity in the symmetry of the respective wavefunctions. For
metallic tubes, we observe the splittings of the first VHS and the
occurrence of a pseudogap. As a
consequence, the wavefunction symmetry breaking remains, which is
demonstrated by conductance images showing strong symmetry variations
within a few meV. Symmetry analysis of the wavefunctions within
an analytical tight-binding approach and calculations of STM
images using density-functional based calculations in the
Tersoff-Hamann formalism give a complete theoretical description of these findings.

\begin{figure*}[!ht]
\centering
 \includegraphics[width=1\textwidth]{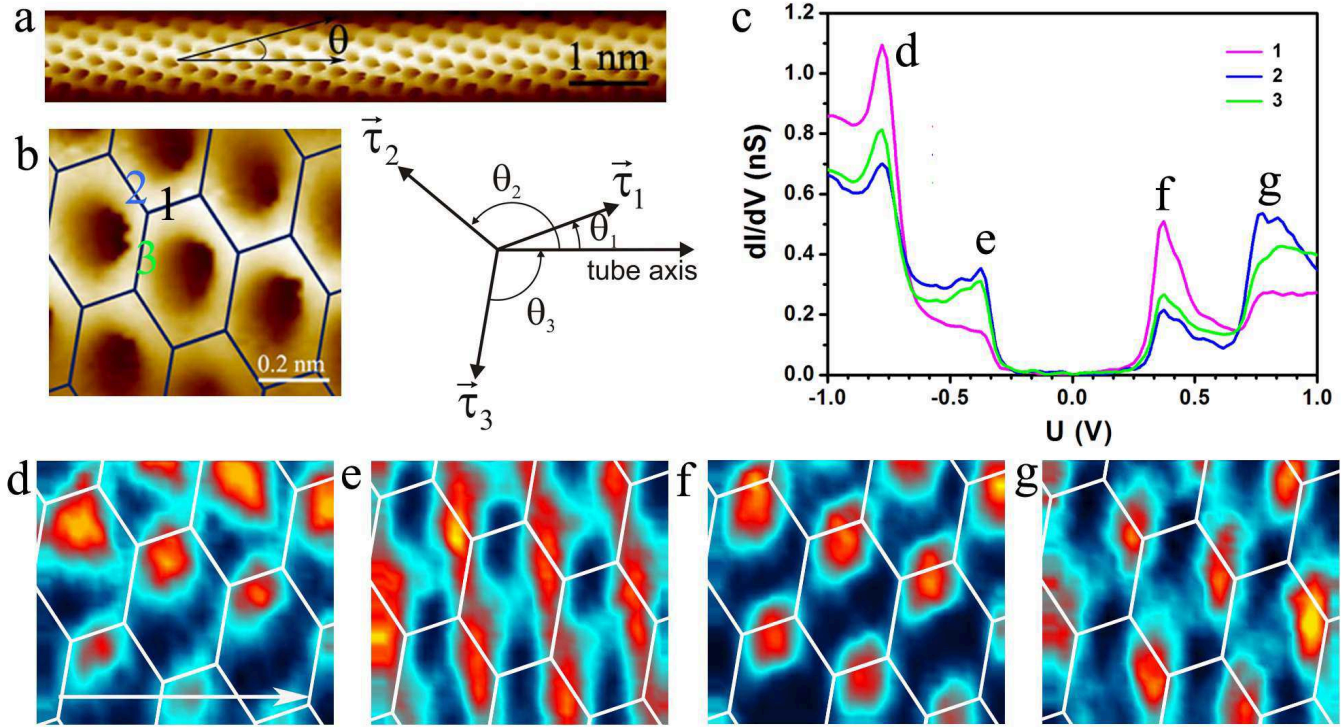}
   \caption{(a) large scale constant
   current topographic image of a $(15,5)$ semiconducting tube. (b) topographic image at atomic scale recorded
   during the CITS measurement with labelling of the three different
   bonds (V$_{bias}$ = 1 V, I = 0.2 nA). The scheme indicates notations used in the tight-binding discussion. $\vec{\tau}{_\alpha}$ and
   ${\theta}{_\alpha}$ are the three C-C bonds vectors and their angles with the tube axis respectively.
   (c) STS measured at three different C-C bonds. 
   (d)-(g): Energy-dependent conductance images of the 
    tube at -0.78 , -0.37 , 0.37, and 0.78 V
   respectively.  The arrow indicates the tube axis.}
\label{semicon}
\end{figure*}

\paragraph{Semiconducting tubes}
We present first a detailed study of a semiconducting tube. From the
topographic image (Fig.\ref{semicon}a), we estimate a chiral
angle $\theta$ about 15$\pm$1$^{\circ}$ and a diameter of
1.4$\pm$0.1 nm, leading to (15,5) chiral indices for the
studied nanotube. The $dI/dV$ spectra measured on the three different
C-C bond orientations (Fig.\ref{semicon}b) exhibit four peaks at -0.78, -0.37, +0.37,
and +0.78 V, corresponding to four VHS denoted
VHS-2, VHS-1, VHS+1 and VHS+2, respectively (Fig.\ref{semicon}c). The wavefunctions at the corresponding energies will be denoted HOMO-1, HOMO, LUMO and LUMO+1 respectively.
The most striking feature of the spectra is
that the relative intensity of the VHS peaks is
position-dependent. Indeed, comparing the spectra measured above
bond 1 and bond 2, the four peaks vary in an alternate way,
\textit{i.e.} VHS-2 decreases while VHS-1 increases, VHS+1 decreases
and VHS+2 increases. A similar variation is also observed between bond
2 and bond 3 but with a smaller amplitude.

Fig.\ref{semicon}d-g show energy-dependent differential
conductance images ($dI/dV$ maps) of the semiconducting tube. The
significant deformation of the hexagonal lattice is mostly due to the drift occuring during the
long acquisition time needed for current imaging tunneling spectroscopy (CITS) measurements as well as
distorsion induced by the curvature of the nanotube
\cite{Meunier1998}.  By contrast to the topography image which is energy
integrated, $dI/dV$ maps reflect the LDOS at a given energy. It clearly displays C-C bond asymmetry, while
keeping the translational invariance of the lattice. Alternating image
types are observed when going from HOMO-1 to LUMO+1 : at -0.78 V, the
intensity is maximum above bonds of type 1, while at -0.37 V there is
a node above bond 1 and maxima above bond 2 and 3. At +0.37 V the
pattern is similar to HOMO-1 while at +0.78 V it is similar to
HOMO. These images demonstrate the complementary nature of the wavefunctions at the VHS energies, which is completely consistent with the
asymmetries of the spectra shown in Fig.\ \ref{semicon}c. We stress
that the present measurements were performed far from the extremity of
the tube and that the topographic image presents a hexagonal network
with the same height for the 3 bond orientations and without any evidence
of the presence of defects. Such alternating patterns have been
observed systematically on the semiconducting tubes, either
individually lying on the substrate or assembled in bundles.

\begin{figure}[!ht]
\centering
 \includegraphics[width=.7\linewidth]{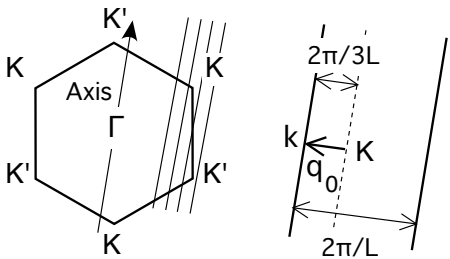}
   \caption{Brillouin zone of the hexagonal graphene lattice (left)
   with a close-up view at the $K$ point (right). The
   parallel lines are the cutting lines of allowed $\vec{k}$ for a 
   $n-m=3p+1$ semiconducting nanotube. The arrow indicates the tube axis direction.}.
\label{BZ}
\end{figure}

Our observation are consistent with the general symmetry arguments in the reciprocal space put forward by Kane and Mele \cite{Kane1999}.
A chemist's approach in
real space appears to be very fruitful also as we show now.
We consider first the VHS of a semiconducting tube. The corresponding Bloch eigenstates have only  a double trivial degeneracy $\pm\vec{k}$, and $\vec{k}$ lies in the vicinity of the $K$ point of the Brillouin zone of the graphene sheet \cite{NT1}. Indeed, the rolling up of the sheet implies that $\vec{k}$ should be on the so-called cutting lines parallel to the axis of the tube and separated by $2\pi/L$, where $L$ is the length of the chiral vector, equal to $\pi d$, $d$ being the diameter of the tube. If $(n,m)$ are the usual coordinates of the chiral vector, the position of the cutting lines with respect to the $K$ point depends on the value of $n-m$ modulo 3. In the case of a semiconducting tube, $n-m=3p \pm 1$, and  the nearest line is at a distance $2\pi/3L$ on the left (resp. right) hand side of $K$ when $n-m=3p+1$ (resp. 3p-1); see Fig.\ \ref{BZ}. Close to this point, the energy is proportional to $|\vec k-\vec{K}|$ and  the LUMO state when $n-m=3p+1$ corresponds to $\vec{k}=\vec{K}+\vec{q_0}$ where $\vec{q_0}$ is a vector of length $2\pi/3L$ pointing along the direction of the chiral vector in the negative direction. The other $-\vec{k}$ value is close to the $K'$ point of the Brillouin zone. 

The eigenstates $\ket{\psi_{\vec{k}}}$  are built from two independent Bloch states $\ket{\psi_{\vec{k}}}=C_1\ket{\vec{u_1}}+C_2\ket{\vec{u_2}}$. These two Bloch states are linear combinations of atomic orbitals on the two sublattices 1 and 2 of the graphene structure, $\braket{\vec{r}}{\vec{u}_{1(2)}}=\sum_{\vec{r_n}\in 1(2)}e^{i\vec{k}.\vec{r_n}}\phi(\vec{r}-\vec{r_n})$, where $\phi(\vec{r}-\vec{r_n})$ is the $\pi$ atomic orbital centred on site $\vec{r_n}$. %
It turns out that the ratio $C_1/C_2$ just introduces a phase factor related to the direction of the $\vec{q}_0$ vector, \textit{i.e.\ }finally to the chiral angle $\theta$, $C_1/C_2=\text{sgn}(E)\exp i(\theta+\varphi)$, where  $\varphi$  depends on  the notations used. 
For tubes such that $n-m=3p-1$, $\vec{q_0}$ is in the opposite direction and the sign of $C_1/C_2$ changes. We calculate now the electronic density $\rho(\vec{r})=|\braket{\vec{r}}{\psi_{\pm\vec{k}}}|^2 $. This involves an expansion in products of overlap functions $\phi(\vec{r}-\vec{r_n})\phi(\vec{r}-\vec{r_m})$. Keeping the main contributions when $\vec{r_n}-\vec{r_m}$ either vanishes or is equal to a nearest neighbour vector, we obtain an expression for the  local density of states $n(\vec{r},E)$, which is the quantity measured in STS. $n(\vec{r},E)$ can be written as $n(E) \times (\rho_{\text{at}}(\vec{r})+\rho_{\text{int}}(\vec{r}))$, where $n(E)$ is the density of states, $\rho_{\text{at}}(\vec{r})$ is the superposition of electronic atomic densities and $\rho_{\text{int}}(\vec{r})$ is the interference term responsible for the breaking of the sixfold symmetry:
\begin{equation*}
\rho_{\text{int}}(\vec{r})\simeq 2 \phi(\vec{r})\phi(\vec{r}-\vec{\tau}{_\alpha}
)\times \text{sgn}(E) \cos \theta_\alpha,
\label{rhoint}
\end{equation*}
where $\theta_\alpha$ is the angle between the tube axis and the bonds $\vec{\tau}_{\alpha}: \theta_\alpha=\theta$ or $\theta\pm2\pi/3$, the chiral angle $\theta$ being here precisely defined as the smallest angle between a bond and the tube axis, $0\leq \theta \leq \pi/6$. 
\begin{figure*}[!ht]
\centering
 \includegraphics[width=1\textwidth]{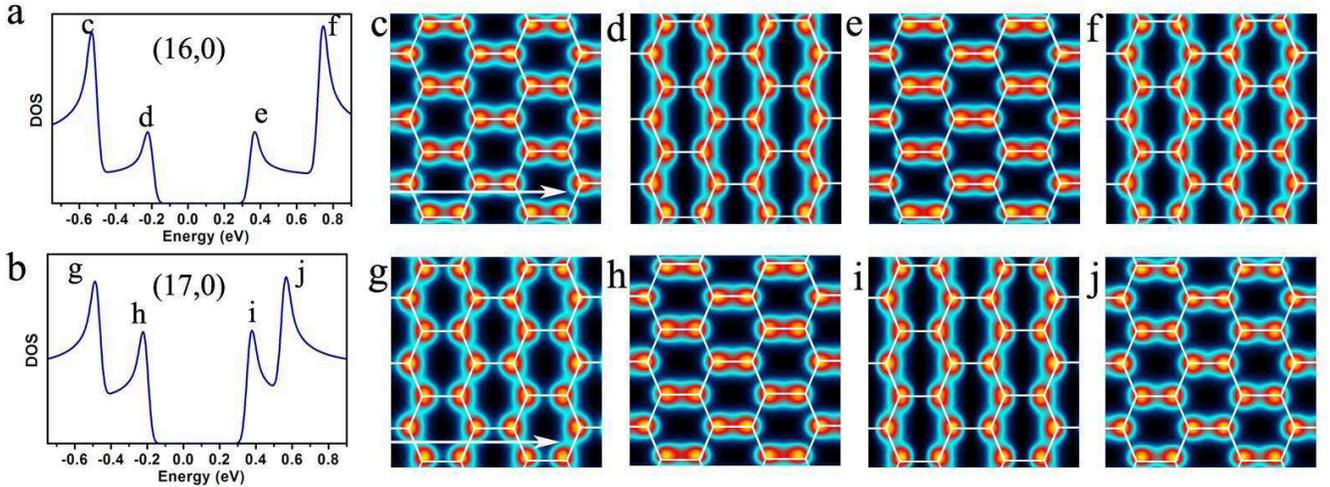}
   \caption{(a) Total DOS of semiconducting (16,0) and (17,0)
   tubes. (c) to (f) show a simulated STM conductance images at
   VHS-2 (-0.52 eV), VHS-1 (-0.21 eV), VHS+1 (0.35 eV) and VHS+2 (0.73 eV) for the (16,0) tube.
   (g) to (j) show a simulated STM conductance images at
   VHS-2 (-0.47 eV), VHS-1 (-0.21 eV), VHS+1 (0.36 eV) and VHS+2 (0.55 eV) for the (17,0) tube.}
\label{DFTsemicon}
\end{figure*}
Because of the dependence on the sign of $E$, the HOMO and LUMO contributions are complementary. In the case of the second VHS, the $\vec{q}_0$ vector should be replaced by $-2\vec{q}_0$ which shows that the singular contribution at the second VHS are also complementary of those of the first ones. Finally images of tubes where $n-m=3p-1$ are also complementary of those where $n-m=3p+1$ since here the $\vec{q}_0$ vector shoud be replaced by $-\vec{q}_0$. Basically two types of images are therefore expected at the singularities depending on the sign in front of $\cos\theta_\alpha$ in Eq.\ (\ref{rhoint}). We call images of type I the images corresponding to a positive sign. In this case the density above the bonds pointing close to the tube axis (angle $\theta$) is reinforced, all the more when $\theta$ is small, \textit{i.e.} close to a zig-zag orientation $(\cos\theta\simeq 1)$. The complementary image (image of type II) presents stripe reinforcements on the bonds perpendicular to the axis $(-\cos\theta_\alpha\simeq 1/2)$. Close to an armchair orientation, the complementarity concerns principally the bonds close to the axis $(\cos\theta_\alpha\simeq\pm\sqrt{3}/2)$, the third one being unaffected $(\cos\theta_\alpha\simeq 0)$. 
As can be seen in Fig.\  \ref{semicon} our experimental results are completely typical of the behaviour when $\cos\theta$ is close to one. This is consistent with our assignement (15,5) for the chiral indices: image of type I for the LUMO state  when $n-m=3p+1$ \cite{Kane1999}.

\begin{figure*}[!ht]
\centering
\includegraphics[width=0.95\textwidth]{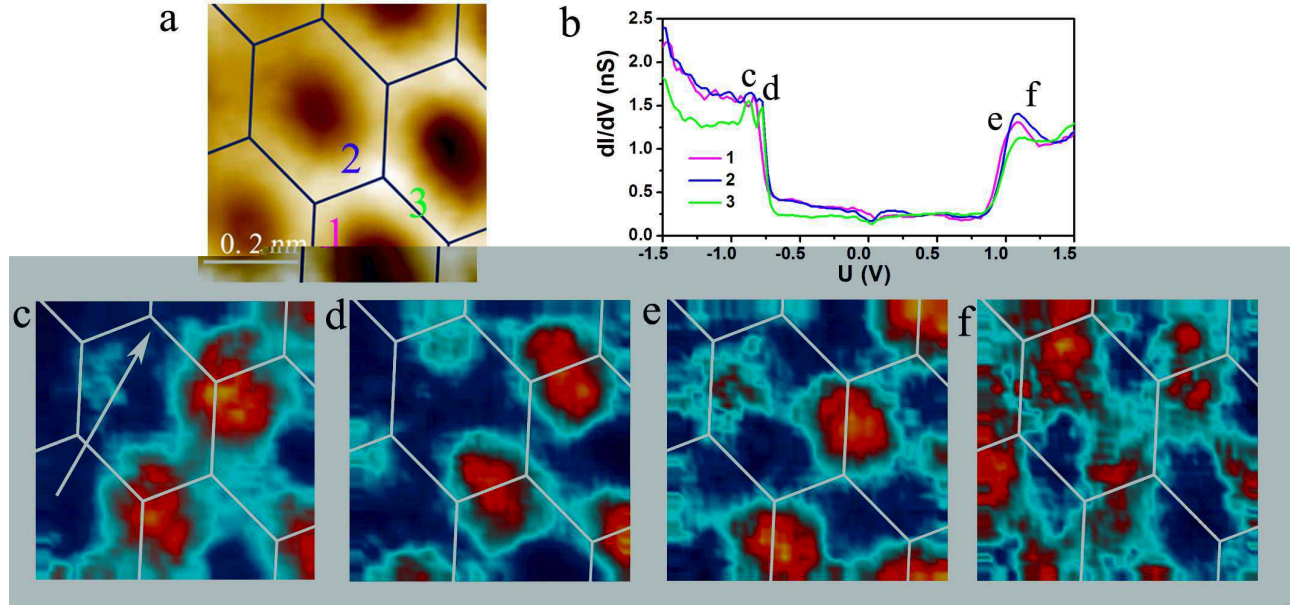}
   \caption{Topographic image (a) of a metallic carbon nanotube ($\theta=26\pm1^{\circ}$) recorded during the CITS measurement with labelling of the 3 bonds where STS point spectra were measured (b). (c)-(f): Energy-dependent conductance images of a metallic
   tube at -0.81, -0.73, 0.96, and 1.10 V respectively.
   The arrow indicates the tube axis.
   }
   \label{metallic1}
\end{figure*}

\begin{figure*}[!htbp] \centering
\includegraphics[width=1\textwidth]{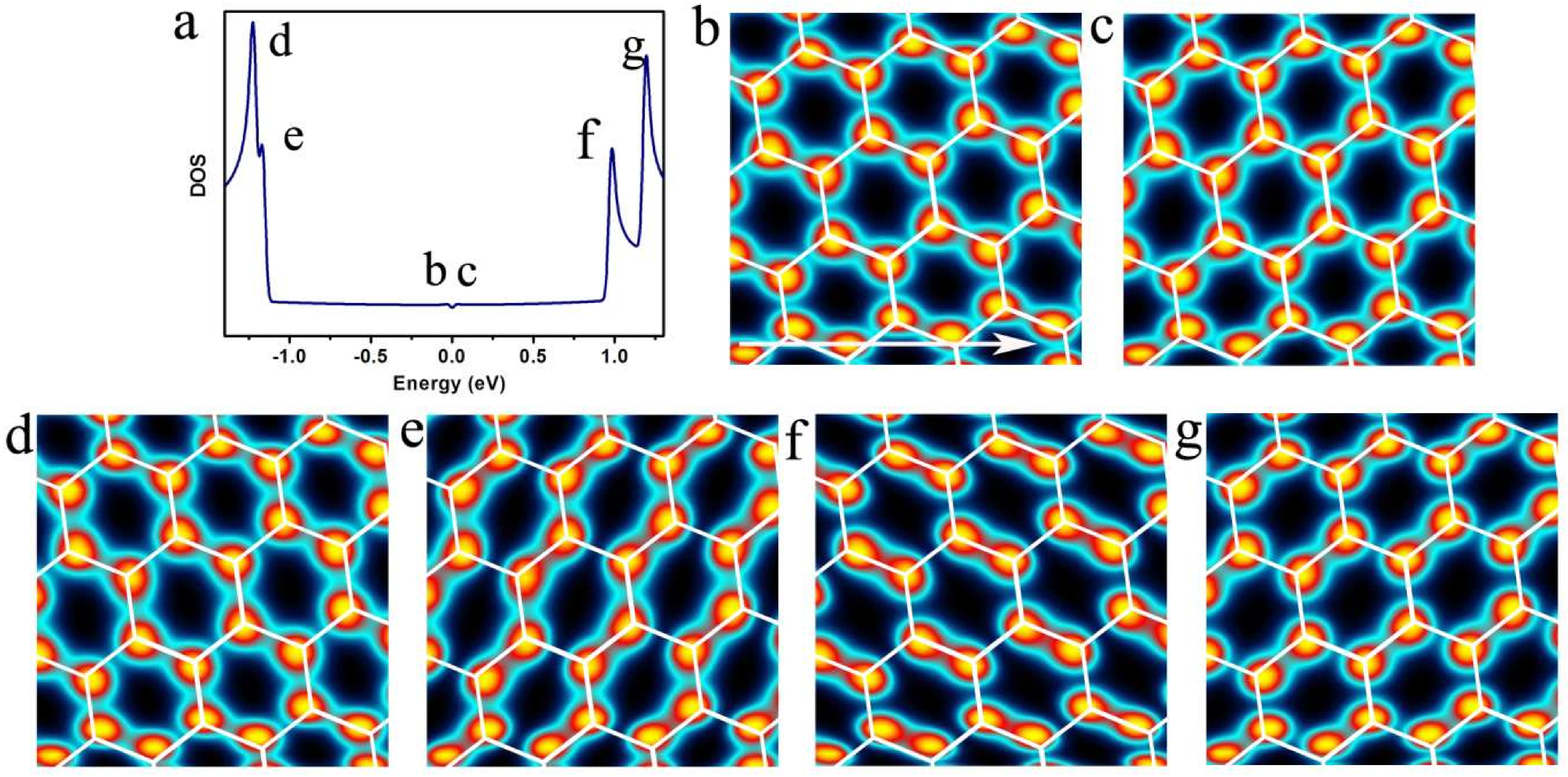}
   \caption{(a) DOS of a (8,5) tube as calculated by DFT methods. (b) and (c) Quantum conductance image at the negative and positive edge of the pseudogap, respectively. (d-g) Quantum conductance at VHS-2, VHS-1, VHS+1 and VHS+2.}
   \label{85}
\end{figure*}

This ``striping effect'' has been simulated in STM images for semiconducting
nanotubes at bias corresponding to VHS $\pm$ 1 within a tight binding approach \cite{meunier2004}.
To validate further our analysis, we have performed first-principles calculations to simulate STM images
on semiconducting zigzag (16,0) and (17,0) nanotubes. The DOS are reproduced
on Fig. \ref{DFTsemicon}a-b. Fig \ref{DFTsemicon}c-j show the simulated conductance
images at energies of VHS-2, VHS-1, VHS+1 and VHS+2 for successively the (16,0) and (17,0) nanotubes. STM
simulations reproduce the ``striping effect'' experimentally observed. (16,0) is a $n-m=3p+1$ tube and presents
a type II behaviour for HOMO and a type I for LUMO in total agreement with measurements and tight-binding prediction on the 3p+1 tubes. Simulations on (17,0) which is a $n-m=3p-1$ tube exhibit a complementary behaviour i.e. a type I for HOMO and a type II for LUMO. These simulations demonstrate the validity of the tight-binding based interpretation
presented above.

\paragraph{Metallic tubes}
We present now STS results on a metallic tube. Fig.\ \ref{metallic1}
shows spectra and $dI/dV$ maps associated with the first
VHS. Remarkably, two kinds of image types are observed around each
VHS. From the tight-binding model previously described, the VHS of metallic tubes
correspond to the four states $\ket{\pm\vec{K}\pm 3\vec{q}_0}$ instead of the doubled degeneracy previously mentioned for semiconducting tubes. An analysis of wavefunction symmetries, similar to the one presented before, do not predict the two observed complementary wavefunctions. However, because of the so-called trigonal warping effect
the degeneracy between the states corresponding to opposite values of
$3\vec{q}_0$ is lifted (except for armchair nanotubes) as observed
experimentally and reproduced in calculations \cite{Saito2000,
Reich2000}.  We then expect, as for semiconducting nanotubes,
complementary images associated with the splitted VHS. This is clearly visible
in experimental conductance spectra. Fig. \ref{metallic1}c-f show
indeed that bond 1 is reinforced at -0.81 V and bond 2 at -0.73
V. Bond 3, almost perpendicular to the axis, is less visible in both
cases, as predicted for a tube close to an armchair orientation,
($\theta=26\pm1^{\circ}$). At positive bias, on the left hand side of
the peak at 0.96 V, bond 1 is bright in the $dI/dV$ map (Fig.\
\ref{metallic1}e) as expected (image of type I). The conductance image
shown in Fig. \ref{metallic1}f is more complex due to the limited
experimental resolution, but it exhibits brightest spots localized on bond 2 instead of bond 1 in Fig.\
\ref{metallic1}e. This is again in line with the tight binding description.

The observed complementary sequence can be shown
to be consistent with the sign of the warping effect: image of type I
for the lowest VHS at positive energy. To confirm this
effect on splitted VHS of metallic tubes, we performed 
first principles simulations on a
(8,5) nanotube. The calculated DOS reported in Fig. \ref{85}a nicely reproduces
the splitting of the VHS and also exhibit a dip around the Fermi level which will be discussed later.

The calculated conductance images at the energies of the 4 peaks represented in Fig. \ref{85}d-g show
alternation of images of typeI and II confirming the broken sixfold symmetry
on metallic tubes. This is totally consistent with the observation and tight binding prediction of striping effect at
split VHS of metallic tubes.

\begin{figure}[!htbp] \centering
\includegraphics[width=1\linewidth]{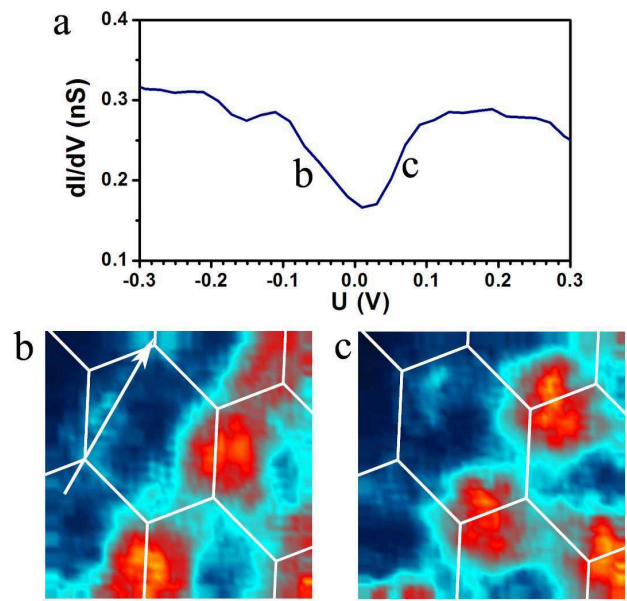}
   \caption{(top) Zoom of the metallic STS around the Fermi level.
   (a)-(b): Energy-dependent conductance images of a metallic
   tube around the Fermi-level at - 30 and + 70 mV. The arrow indicates the tube axis.}
   \label{metallic2}
\end{figure}

We conclude with a significant observation of the symmetry breaking
at the band edge of the pseudogap.  The LDOS in
(Fig. \ref{metallic2}(top)) measured by STS clearly shows a dip around the Fermi level corresponding to the pseudogap.
The dI/dV maps recorded at -30 mV (Fig. \ref{metallic2}a) and +70 mV (Fig. \ref{metallic2}b) exhibit complementary symmetry, type I at negative bias and type II at positive bias.
Indeed, this small gap is characterized by a small $\vec{q}$ perpendicular to the tube
axis. Complementary images of type I and II are therefore also
predicted in the tight-binding model at the lower and upper band edges, respectively,
in nice agreement with our observations . The first principles of the (8,5) tube also confirms the striping effect
at the pseudogap as shown in Fig. \ref{85}b-c. The image at positive bias is complementary of
the image at negative bias. Moreover it is complementary of the image calculated at the first split VHS peak at positive bias (Fig. \ref{85}f) confirming the alternation of symmetry observed experimentally.

To summarize, we have shown that STS allows us to image directly the
wavefunctions of carbon nanotubes associated with band edges and VHS. The wavefunctions
intensities present patterns with broken sixfold symmetry. The images alternate systematically between
only two types when going from one singularity to another. Considering the HOMO orbital, for a
$3p-1$ tube the electron density is concentrated on the C-C having the smaller angle with the tube axis. For a $3p+1$ 
electron density is maximum on the two other bonds. For a metallic tube (3p) the sixfold symmetry is expected to be recovered. However the trigonal warping lifts the degeneracy and pairs of complementary orbitals are revealed around the energy position of the VHS. These results give a complete view of nanotubes molecular orbitals and allow to predict easily any wavefunction of any nanotube with a honeycomb lattice. This global picture is expected to provide essential insight for all cases where the symmetry relation between nanotubes orbitals and their surrounding is a key factor such as functionalization, transport devices or nanotubes sorting.

\section{Experimental and theoretical details}
\paragraph{Sample preparation and STM setup}
STM measurements were performed with a low temperature STM operating at 5K under UHV conditions
(less than $10^{-10}$ mbar). Spectroscopy was achieved in the current
imaging tunneling spectroscopy (CITS) mode. Local $dI/dV$ spectra were
measured with a lock-in amplifier at each point of the
images. Conductance images as well as local point spectra were then
extracted from the data. The SWNTs were synthesized in
a vertical flow aerosol reactor and deposited \emph{in-situ} onto
commercial gold on
 glass surface previously flashed by butane flame in air. After the
 synthesis, the samples were introduced into the UHV system and heated
 to 150$^{\circ}$C during 30 min, for degasing. All measurements were
 performed with tungsten tips.
 \paragraph{DFT and STM simulations}
Density-functional calculations were performed using the SIESTA code within the local density approximation as parametrized by Perdew and Zunger~\cite{PZ}. The core electrons were replaced by nonlocal norm-conserving pseudopotentials~\cite{PSP}.
A double-$\zeta$ basis set of localized atomic orbitals were used for the valence electrons. STM topological images were calculated according to the
Tersoff-Hamann approximation~\cite{Tersoff}.

\section{acknowledgments}
  The authors are indebted to Ph. Lambin for helpfull discussion and
to T. Susi and E. Kauppinen in the NanoMaterials Group of Helsinki
University of Technology for providing the nanotube samples. This
study has been supported by the European Contract STREP ``BCN
Nanotubes'' 30007654-OTP25763, a grant of CNano IdF ``SAMBA'' and the
ANR project ``CEDONA". Simulations have been performed on the
ISCF center (University of Namur) supported by the FRS-FNRS. L.H. is supported by the FRS-FNRS.

\end{document}